\documentclass[sn-basic,Numbered,iicol]{sn-jnl}  

\usepackage{graphicx}
\usepackage{textcomp}
\usepackage{color,xcolor}
\usepackage{caption}

\usepackage{subfigure}
\usepackage{booktabs}
\usepackage{hyperref}
\usepackage[switch]{lineno}
\usepackage{soul}
\usepackage{utfsym}
\usepackage{indentfirst}
\usepackage{times}
\usepackage{latexsym}

\usepackage{graphicx}%
\usepackage{multirow}%
\usepackage{amsmath,amssymb,amsfonts}%
\usepackage{amsthm}%
\usepackage{mathrsfs}%
\usepackage[title]{appendix}%
\usepackage{xcolor}%
\usepackage{textcomp}%
\usepackage{manyfoot}%
\usepackage{booktabs}%
\usepackage{algorithm}%
\usepackage{algorithmicx}%
\usepackage{algpseudocode}%
\usepackage{listings}%

\newtheorem{theorem}{Theorem}
\newtheorem{proposition}[theorem]{Proposition}%
\newtheorem{definition}{Definition}%

\begin{document}

\title[Article Title]{Improving Global Parameter-sharing in Physically Heterogeneous Multi-agent Reinforcement Learning with Unified Action Space}


\author[1,2]{\fnm{Xiaoyang} \sur{Yu}}\email{xiaoyang.yu@bjtu.edu.cn}
\author[1,2]{\fnm{Youfang} \sur{Lin}}
\author[1,2]{\fnm{Shuo} \sur{Wang}}
\author[1,2]{\fnm{Kai} \sur{Lv}}
\author*[1,2]{\fnm{Sheng} \sur{Han}}\email{shhan@bjtu.edu.cn}

\affil[1]{\orgdiv{School of Computer Science and Technology}, \orgname{Beijing Jiaotong University}, \orgaddress{\city{Beijing}, \country{China}}}

\affil[2]{\orgdiv{Beijing Key Laboratory of Traffic Data Analysis and Mining}, \orgname{Beijing Jiaotong University}, \orgaddress{\city{Beijing}, \country{China}}}



\abstract{In a multi-agent system (MAS), action semantics indicates the different influences of agents' actions toward other entities, and can be used to divide agents into groups in a physically heterogeneous MAS. Previous multi-agent reinforcement learning (MARL) algorithms apply global parameter-sharing across different types of heterogeneous agents without careful discrimination of different action semantics. This common implementation decreases the cooperation and coordination between agents in complex situations. However, fully independent agent parameters dramatically increase the computational cost and training difficulty. In order to benefit from the usage of different action semantics while also maintaining a proper parameter-sharing structure, we introduce the Unified Action Space (UAS) to fulfill the requirement. The UAS is the union set of all agent actions with different semantics. All agents first calculate their unified representation in the UAS, and then generate their heterogeneous action policies using different available-action-masks. To further improve the training of extra UAS parameters, we introduce a Cross-Group Inverse (CGI) loss to predict other groups' agent policies with the trajectory information. As a universal method for solving the physically heterogeneous MARL problem, we implement the UAS adding to both value-based and policy-based MARL algorithms, and propose two practical algorithms: U-QMIX and U-MAPPO. Experimental results in the SMAC environment prove the effectiveness of both U-QMIX and U-MAPPO compared with several state-of-the-art MARL methods.}




\maketitle

\section{Introduction}
The researches on multi-agent system (MAS) and multi-agent reinforcement learning (MARL) have been remarkably expanding \cite{APPI-oroojlooy2022review,survey+gronauer2022multi,zhu2024survey,hernandez2019survey} and have been applied to solve complex problems in the industry, such as traffic management of autonomous drones \cite{balazs2023decentralized}, multi-agent air combat \cite{NCA-sun2023multi}, power control in wireless networks \cite{amorosa2023multi}, battery-swapping system for Unmanned Ground Vehicles (UGV) \cite{holand2024battery}, resource allocation problem \cite{allahham2022multi,NCA-lu2023maddpg}, portfolio management \cite{NCA-ma2023multi}, dialogue state tracking \cite{NCA-huang2024mgcrl}, and traffic lights control problem \cite{traffic+liu2022distributed}. Previous deep MARL algorithms have achieved impressive results in cooperative MARL environment \cite{survey+gronauer2022multi,NCA-zhao2024mimic}. The Starcraft Multi-Agent Challenges (SMAC) environment \cite{smac+samvelyan2019starcraft} is a popular yet powerful research environment. \par
The SMAC environment is a multi-agent real-time tactical micromanagement game. The basic mission scenario is that two adversarial MAS fight against each other. The goal is to train a MARL algorithm to win enemy agents controlled by the internal script of the SMAC environment. The algorithm requires learning tactics and skills for choosing the best action response while utilizing the properties of different agents. \par
Heterogeneous MAS is very common in real-world scenarios \cite{clauset2009power,wang2024what,wang2023skill,lv2023spatially,lv2020pose}, such as wireless network accessibility problem \cite{wifi+yu2021multi}, heterogeneous LLM agents for financial sentiment analysis \cite{xing2024designing} and multi-agent robotic systems \cite{robotic+ivic2020motion,robotic+yoon2019learning}. Traditionally, heterogeneity of MAS can be categorized into two classes: \textit{Physical} (or, \textit{Morphological}) and \textit{Behavioral} \cite{Def-HetGPPO-bettini2023heterogeneous,Def-wilson2022evolution,Def-wilson2022performance}. \textit{Physical} heterogeneity denotes the difference of MAS components in hardware, physical constraints, or characteristics, indicating their different capabilities and goals. On the other hand, \textit{behavioral} heterogeneity denotes the difference of MAS components in software, behavioral models, or dynamics. In \cite{GHQ-yu2024ghq}, the authors give a formal definition of physical heterogeneity from the perspective of \textit{local transition function} and \textit{ideal action object}. Previous mainstream approaches use policy-based actor-critic algorithms to solve the physically heterogeneous MARL problem simply with various individual agent policies \cite{happo+kuba2021trust,mapg+bono2018cooperative}. \par
\textit{Global parameter-sharing} is a common yet efficient implementation method in both value-based and policy-based MARL algorithms, such as \cite{rashid2020+qmix-jmlr,mappo+yu2021surprising}. It is one of the simplest and the most powerful ways to fulfill a \textit{cooperative} yet \textit{scalable} MARL algorithm. In \cite{happo+kuba2021trust}, the authors point out that global parameter-sharing may lead to exponentially bad local optimal policy. In this paper, we further conclude that global parameter-sharing may also lead to extremely bad local optimality in the physically heterogeneous MARL problem without properly handling different action semantics. However, as is demonstrated in \cite{homophily+dong2021birds}, homophily is necessary for cooperative MAS to achieve a better joint cooperating policy. For example, in human teamwork cooperation, good collaboration is more likely achieved through a ``tacit'' joint action policy. Therefore, the best way to deal with this dilemma is to realize a proper global parameter-sharing structure capable of attaining good cooperation and learning efficiency while avoiding bad local optimality. \par
As is described in \cite{ASN-wang2019action}, the term \textit{action semantics} describes the property that agent actions have different influences towards other entities. The authors then divide the agent action set \textit{A} into two subsets: $A_{in}$ and $A_{out}$. $A_{in}$ includes the actions that only affect the environment or the acting agent itself. On the contrary, $A_{out}$ consists of the actions that directly influence other agents or enemies. In this paper, we divide the $A_{out}$ into two sub-groups concerning their different action object. Furthermore, we point out that the existence of different action semantics results in physical heterogeneity since different action semantics indicate different \textit{ideal action objects (IO)}.\par
In this paper, we propose the Unified Action Space (UAS) and the U-MAPPO and U-QMIX algorithms for implementing a better global parameter-sharing structure in physically heterogeneous MARL problems.
First, we give an example to prove that global parameter-sharing may lead to extremely bad local optimality in the physically heterogeneous MARL problem without properly handling different action semantics, which provides the necessity and motivation of our research.
Second, we introduce the Unified Action Space (UAS), which is the union set of all agent actions with different action semantics. Heterogeneous agents can be distinguished through different available-action-masks (AM). All agents first calculate their unified representation in the UAS, and then generate their heterogeneous action policies using different AMs.
Third, in order to further improve the training of UAS parameters, we introduce a Cross-Group Inverse (CGI) loss to predict other groups' agent policies with the group's trajectory information.
Finally, we combine the UAS with a policy-based algorithm (MAPPO, \cite{mappo+yu2021surprising}) and a value-based algorithm (QMIX, \cite{rashid2020+qmix-jmlr}) as implementing examples to prove the effectiveness and universal of our UAS method. U-MAPPO and U-QMIX successfully leverage the advantages of parameter-sharing while also reaching a high winning rate (WR). Results in the SMAC environment prove the effectiveness of the two proposed algorithms compared with several state-of-the-art MARL methods.
\par
The rest of the content is as follows: we summarize some related works in section \ref{Related Works}; we give the definitions and the example of a bad parameter-sharing structure in section \ref{Definition and Analysis}; we provide details about the UAS methods and the U-MAPPO and U-QMIX algorithms in section \ref{Method}; we present detailed environmental design and discuss results of our experiments in section \ref{Experiments and Results}; and finally we draw some conclusion in section \ref{Conclusion}.

\section{Related Works}\label{Related Works}
\subsection{Multi-agent Reinforcement Learning and Heterogeneity}
The centralized training with decentralized execution (CTDE) paradigm \cite{ctde+foerster2016learning,ctde+kraemer2016multi,ctde+gupta2017cooperative}, which requests agents not to use state $S$ during execution, has become one of the most successful MARL paradigms in SMAC environment. \par
For policy-based methods, a simple implementation of CTDE is the ``centralized critic and decentralized actor'' (CCDA) structure. The state $S$ is only available for the critic network during the centralized training phase. COMA \cite{foerster2018+coma} uses a counter-factural baseline for learning a better policy. MAPPO \cite{mappo+yu2021surprising} directly implements the CCDA structure with a global critic and a shared actor. HAPPO \cite{happo+kuba2021trust} introduces the ``multi-agent advantage decomposition'' (MAAD) lemma and designs a sequential executing and updating structure for the independent actors. MAT \cite{MAT-wen2022multi} further develops the HAPPO by replacing the backbone of HAPPO into the Transformer-style network structure, and naturally uses the auto-regressive decoder as the sequential actor.\par
For value-based methods, the mainstream is the value factorization method, whose goal is to learn a centralized and factorized joint action-value function $Q_{tot}$ and the factorization structure: $Q_{tot} \to Q_i$. In order to factorize $Q_{tot}$ and use the \textit{argmax} policy of $Q_i$ to select actions, the Individual-Global-Max (IGM) consistency \cite{son2019qtran} is required:
\begin{equation}
\arg \max_{\boldsymbol{a}} Q_{tot}(\boldsymbol{\tau}, s) =
\begin{pmatrix}
  \arg \max_a Q_{1}(\tau_1)\\
  ..., \\
  \arg \max_a Q_k(\tau_k)
\end{pmatrix} \ .
\end{equation}
\par
VDN \cite{vdn+sunehag2017value} simply uses the sum of local $Q_i$ functions to represent $Q_{tot}$. QMIX \cite{rashid2020+qmix-jmlr} changes the factorization function from additivity to monotonicity, which dramatically enrich the embedding ability of the network. RODE \cite{wang2020rode} and ROMA \cite{wang2020roma} are \textit{role-based} algorithms, which learn role policies online. In general, several key hyper-parameters define the clustering and using of role policies. Therefore, the performance of ROMA and RODE highly depends on searching for the optimal hyper-parameters, not to mention that the default training step of ROMA is 20M time-steps. CDS \cite{li2021+CDS} maximizes the mutual information between the agent trajectory and its own agent ID to acquire distinguishable individual local $Q$ functions. ASN \cite{ASN-wang2019action} analyzes different actions’ influence on other agents and uses different sub-networks to capture the difference. UNMAS \cite{UNMAS-chai2023unmas} proposes the self-weighting mixing network and the individual action-value networks to adapt to the changes in agent numbers and action space.\par
Traditionally, the study about heterogeneity in MARL is insufficient, because heterogeneous MARL has been considered as a special case of homogeneous MARL. HAPPO lacks specific analysis and sufficient experiments to analyze the effect of heterogeneity. Its successors, MAT and HASAC \cite{HASAC-2024liumaximum}, only modify the network structure and optimizing methods and still lack the study in heterogeneity. SHPPO \cite{SHPPO-guo2024heterogeneous} uses a latent network to learn embedding variables for heterogeneous decision-making. In other field of MAS, FedMRL \cite{FedMRL-sahoo2024fedmrl} develops a novel federated multi-agent deep reinforcement learning framework to address data heterogeneity for medical imaging. HeR-DRL \cite{HeR-DRL-zhou2024her}.\par
Recently, HetGPPO \cite{Def-HetGPPO-bettini2023heterogeneous} conducts a taxonomical study on heterogeneity classification, and introduces a graph neural network heterogeneous agent-wise communication. GHQ \cite{GHQ-yu2024ghq} gives a formal definition of the \textit{local transition} heterogeneity, which is equal to the \textit{physical} heterogeneity, and uses a grouping method and hybrid value factorization structure to learn grouped Q functions for choosing actions.

\subsection{Global parameter-sharing}
The global parameter-sharing mechanism is a common implementation widely used in MARL. If all agents are homogeneous, their policy or Q networks can be trained more efficiently with less computing complexity by adopting parameter-sharing \cite{mappo+yu2021surprising,rashid2020+qmix-jmlr}. In order to deal with heterogeneous agents, in SMAC environment, many algorithms introduce extra information to identify agent ID or agent type. The advantage of global parameter-sharing is quite obvious. Since all agents use a shared network to choose action, the \textit{scalability} problem does not affect the network structure and could be easier to solve. In general, \textit{cooperation} requires necessary similarity and has been proved in \cite{homophily+dong2021birds}.\par
However, in HAPPO \cite{happo+kuba2021trust}, the authors prove that global parameter-sharing may lead to exponentially bad local optimal policy in an XOR-style situation. Therefore, HAPPO uses independent actor networks to choose actions. ASN \cite{ASN-wang2019action} enables parameter-sharing between different sub-modules of the network in the same homogeneous agent group. CDS \cite{li2021+CDS} introduces diversity into the optimization and representation of QMIX to overcome the shortage of global parameter-sharing.

\section{Definition and Analysis}\label{Definition and Analysis}
In this section, our goal is to prove the necessity of considering action semantics while adopting parameter-sharing in MARL. First, We introduce fundamental concepts and definitions in \ref{Preliminaries}. Next, we give a possible example of a bad joint action policy under parameter-sharing without distinguishing different action semantics.

\subsection{Preliminaries}\label{Preliminaries}
In this paper, we study the physical heterogeneous cooperative MARL problems using the decentralized partially observable Markov decision process (Dec-POMDP) \cite{DEC-POMDP+oliehoek2016concise} scheme. The problem is described with a tuple $G=\left\langle S, A, O; P, \Omega, R; \gamma, N, K, T \right\rangle$. $N= \{ 1, ..., n_{i}, ..., n  \} $ denotes the finite set of $n$ agents, $K= \{ 1, ..., k_{i}, ...,  k  \} $ denotes the finite set of $k$ agent groups, and $\gamma \in [0,1)$ is the reward discounting factor. $s \in S$ denotes the true state of the environment with complete information and is not accessible for decentralized components. At each time-step $t \le T$, agent $i \in N$, from group $k_{i} \in K$, receives an individual partial observation $o_i^t$ from the environment and chooses an action $a_i^t \in A_i$ from its local available action set $A_i$. $A_i$ is masked from the global action set $A$ with agent's local available-action-mask $AM_i$. Actions of all agents form a joint action $\boldsymbol{a}^t = (a_1^t, ..., a_n^t) \in A = (A_1, ..., A_n)$ at every time-step. The environment receives the joint action $\boldsymbol{a}^t$ and returns a next-state $s^{t+1}$ according to the transition function $P (s^{t+1}|s^{t}, \boldsymbol{a}^t)$, and a reward $r^t$ according to the global reward function $R(s, \boldsymbol{a}^t)$. All local observations $o_1^t, ..., o_n^t$ are generated from the state $s^t$ according to the observation function $\Omega (s^t, i)$, and form the joint observation $\boldsymbol{o}^t = (o_1^t, ..., o_n^t) \in O$. Observation-action trajectory history $\tau^t_i = \cup_1^t \{ (o^{t-1}_i, a^{t-1}_i)\}$ ($t \ge 1; \tau^0 = o^0$) is the summary of the partial transition tuples of agent $i$ before time-step $t$. Specifically, $\tau_i$ indicates the overall trajectory of agent $i$ through all time-steps $t \le T$. Replay buffer $\mathcal{D} = \cup (\tau, s, r)$ stores all data for batch sampling. Network parameters are notated with $\theta$, $\phi$ and $\psi$.

\subsection{Limitation of Existing Parameter-sharing Methods}\label{Limitation}
As is mentioned in \cite{happo+kuba2021trust}, adopting parameter-sharing among all agents equals adding a constraint $\theta^i = \theta^j, \forall i,j \in N$ to the joint policy space. Here, we point out that utilizing parameter-sharing without properly distinguishing different action semantics can lead to a sub-optimal solution.\par

\begin{proposition}
Considering a fully-cooperative physically heterogeneous MARL problem. Let there be 2 groups of agents, marked as $G_0$ and $G_1$. Both groups have $N$ agents. The actual action space of $G_0$ is $A_0 =  \{ a_0, ..., a_m\}$, while the actual action space of $G_1$ is $A_1 =  \{ a_0, ..., a_m, ..., a_n\}$ ($|a_n| > |a_m| \ge N$). The reward function is defined as:
\begin{equation}
    r=
    \begin{cases}
        \begin{aligned}
            & 1 & (\forall \ n_{id}, \ a_{id} = n_{id}, \ simultaneously) \\
            & 0 & (else)
        \end{aligned}
    \end{cases} \ .
\end{equation}
Let $J^*$ be the optimal joint reward, and $J^*_{\rho}$ be the optimal joint reward under the parameter-sharing constraint. Then we have:
\begin{equation}
    \begin{aligned}
        & \frac{J^*}{J^*_{\rho}} = \frac{1}{(\frac{\rho_r}{N})^{2N}} \ , \\
        & 0< \frac{\rho_r}{N} <1 \ .
    \end{aligned}
\end{equation}
\end{proposition}
\par
\textit{Proof.} Suppose the joint action policy distribution under the parameter-sharing constraint is $\{ \rho_0, ..., \rho_m, ..., \rho_n \}$ for all $2N$ agents in both groups, where $\rho_i$ indicates the probability of action $a_i$. It is obvious that $J^* = 1$, for we can let the joint policy be the deterministic policy that all agents with their id $i \le N$ from both groups choose the same action $a_i$. For $J_{\rho}$, because it is required to simultaneously fulfill $a_{id} = n_{id}$, the expected reward is:
\begin{equation}
    J_{\rho} = \prod_{i=0}^{N} \rho_i^2 \ .
\end{equation}
\par
To maximize $J_{\rho}$, since $\rho_i > 0$ and $N > 0$, we can equivalently maximize $(\prod_{i=0}^{N} \rho_i)^{1/N}$. According to the arithmetic-geometric means inequality, we have:
\begin{equation}
    (\prod_{i=0}^{N} \rho_i)^{1/N} \le \frac{1}{N} \sum_{i=0}^{N} \rho_i \ ,
\end{equation}
where the equality holds if and only if $\forall i,j \le N, \rho_i = \rho_j$. Let $\rho_r = \sum_{i=0}^{N} \rho_i$, and $\forall i \in N, \rho_i = \rho_r / N$. Then we have:
\begin{equation}
    J_{\rho}^* = \prod_{i=0}^{N} (\frac{\rho_r}{N})^2 = (\frac{\rho_r}{N})^{2N}  \ ,
\end{equation}
which is exactly the result. \qed
\par
Through this example, we prove that parameter-sharing can lead to an exponentially bad sub-optimal result as the number of agents increases. In addition, it can be inferred that increasing $\rho_r$ helps to improve $J_{\rho}^*$. However, according to the heterogeneity constraint and the parameter-sharing constraint, the maximum of $\rho_r$ is $\sum_{i=0}^{|a_m|} \rho_i$, which is still smaller than 1. \par

\subsection{Relation between action semantics and physical heterogeneity}\label{Relation-sec3.3}
In this paper, we divide the original agent action set $A$ into three subsets: $A_s$, $A_a$ and $A_e$. $A_s$ indicates $self$ actions that only affect the agent itself, $i.e.$ moving actions. $A_a$ indicates $ally$ actions that only affect the ally agents, $i.e.$ healing actions. $A_e$ indicates $enemy$ actions that only affect the enemy entities, $i.e.$ attacking actions. \par
This classification method follows the different action semantics mentioned in \cite{ASN-wang2019action}, and further distinguishes different action objects. Considering the definition of physical heterogeneity provided in \cite{GHQ-yu2024ghq}, the different \textit{ideal action objects (IO)} is sufficient to determine and analyze the heterogeneity. \par
Therefore, the difference in agents' action semantics results in physical heterogeneity. Additionally, it is beneficial to utilize the different action semantics for better cooperative policy learning in physically heterogeneous MARL problems.

\section{Method}\label{Method}
\subsection{Unified Action Space}\label{UAS}
As is analyzed in section \ref{Limitation}, adopting global parameter-sharing in physically heterogeneous MARL problems without proper structure can lead to harmful sub-optimality. Therefore, we introduce the \textit{Unified Action Space} as a solution. \par
\begin{definition}
Unified Action Space (UAS): Divide all agents' $i \in N$ local available action set $A_i$ into the three sets with different action semantics: $A_s$, $A_a$ and $A_e$. Then, the union set of the three sets of actions forms the Unified Action Space $A^U = A_s \cup A_a \cup A_e$.
\end{definition}
\par
It is obvious that the UAS is a superset to the original action space $A^U \supseteq A$, as the original set is not forced to discriminate different \textit{action semantics}. As is analyzed in section \ref{Relation-sec3.3}, in physically heterogeneous MARL problems, different action semantics determine physical heterogeneity and can be used to divide agents into different groups following the \textit{IOG} requirement \cite{GHQ-yu2024ghq}. \par
\begin{figure}
    \centering
    \includegraphics[width=0.99\columnwidth]{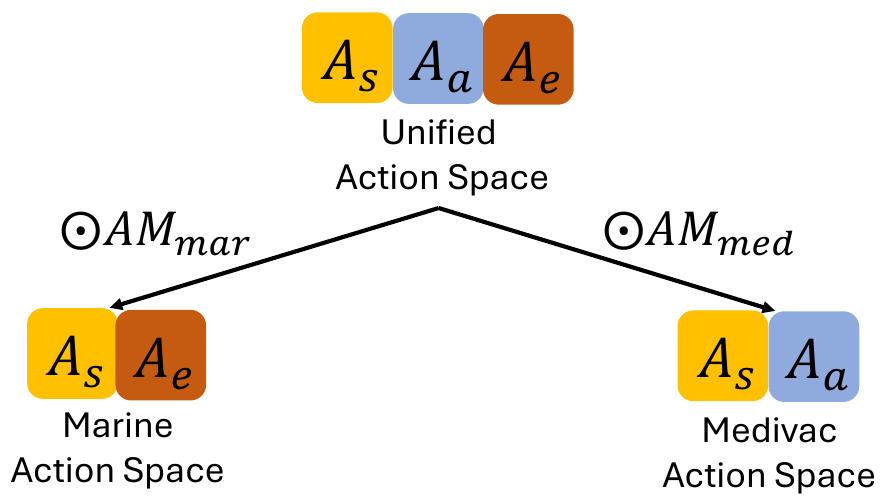}
    \caption{An illustration of the usage of UAS.}
    \label{fig-UAS}
\end{figure}

Furthermore, different action semantics can be identified through different available-action-mask \textit{(AM)} defined on the UAS. For example, in SMAC, \textit{Marine} is an attacking unit whose action semantics include \textit{moving} and \textit{attacking enemies}. As a result, the $AM_{mar}$ only includes two subsets, $A^U \odot AM_{mar} = A_s \cup A_e$. On the other hand, \textit{Medivac} is a supporting unit whose action semantics include \textit{moving} and \textit{healing allies}. Therefore, the $AM_{med}$ is only available for $A^U \odot AM_{med} = A_s \cup A_a$. We point out that the $\odot$ operator indicates the \textit{masking-policy} operation. Fig. \ref{fig-UAS} illustrates the usage of UAS and the different action semantics. \par

\begin{figure*}
\centering
\includegraphics[width=0.99\textwidth]{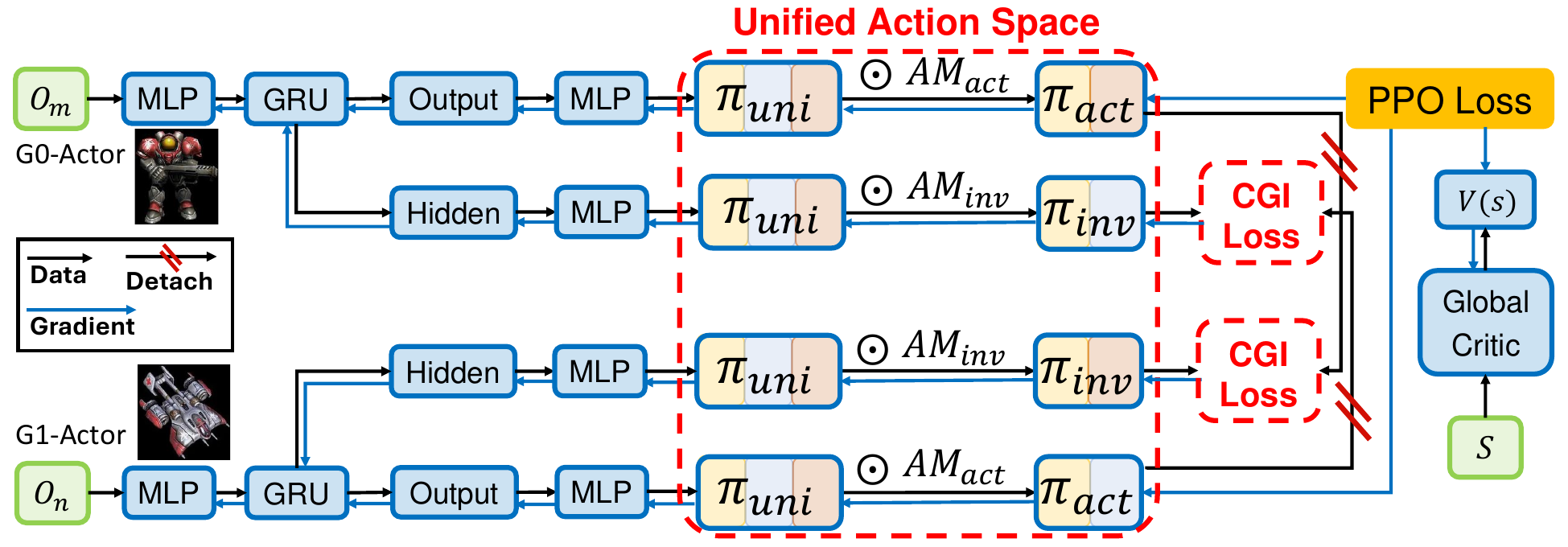}
\caption{An overall framework of U-MAPPO. The overall network consists of three parts: actor network $\boldsymbol{\theta}$, predictor network $\boldsymbol{\psi}$ and critic network $\boldsymbol{\phi}$. The actor network $\boldsymbol{\theta}$ and the predictor network $\boldsymbol{\psi}$ are shared by all agents. They take their corresponding inputs and generate the UAS policies $\boldsymbol{\pi}_{uni}$. Then the $\boldsymbol{\pi}_{uni}$ are masked by different available-action-masks $AM$ to generate the joint action policy $\pi_{\boldsymbol{\theta}}$ and the joint inverse policy $\rho^{inv}_{\boldsymbol{\psi}}$ separately. The critic network $\boldsymbol{\phi}$ is a global network to compute the global value function $V_{\boldsymbol{\phi}}(s)$ during training.}
\label{fig-framework-U-MAPPO}
\end{figure*}

Introducing the UAS allows us to apply global parameter-sharing while also avoiding the limitation discussed in \ref{Limitation}. Considering the example in section \ref{Limitation}, let us suppose that the UAS is $A^U =  \{ a_0, ..., a_n, a_0', ..., a_m'\}$. Then, the UAS policy distribution becomes $\rho^U = \{ \rho_0, ..., \rho_n, \rho_0', ..., \rho_m' \}$ for all agents. For generating group-specific policy, we need to mask the UAS distribution into group policy distribution: $\rho_{G_0} = \rho^U \odot AM_{G_0} = \{ \hat{\rho_0}', ..., \hat{\rho_m}' \}$ and $\rho_{G_1} = \rho^U \odot AM_{G_1} = \{ \hat{\rho_0}, ..., \hat{\rho_n} \}$. Because the mask operation re-normalizes group policies, we have $\sum_{i=0}^{|G_0|} \hat{\rho_i}' = 1$ and $\sum_{i=0}^{|G_1|} \hat{\rho_i} = 1$. Consequently, the policies of the two groups become independent and we can implement the same deterministic global optimal joint policy as $J^*$ requires without restriction. As a result, the $J_{\rho}^*$ with the usage of the UAS and $AM$ becomes 1, which is equivalent to the global maximum $J^*$.

\subsection{Cross-Group Inverse Loss}\label{CGI Loss}
In a multi-agent system (MAS), a cooperative joint policy is usually approached by maximizing the mutual information between agents. However, calculating mutual information must be taken between two groups and the overall computational complexity is proportional to the combination number of 2 groups out of K groups $\Omega (C_K^2)$. \par
Introducing the UAS provides another available way to learn a cooperative policy, which is learning an agent's policy while also predicting other agents' policies. ``Learning with predicting'' is a natural way for humans to practice and perform cooperation, and is useful to develop tacit team policy. To follow the ``learning with predicting'' scheme, we introduce the \textit{Cross-Group Inverse loss (CGI loss)}. \par
The core innovation is the \textit{inverse available-action-mask} $AM^{inv}$. Since the action policy is generated through the masking operation $\rho_i = \rho^U \odot AM_i$, we can generate the action policy of other physically heterogeneous agents by applying their mask to the agent's UAS policy: $\rho_i^{inv} = \hat{\rho}^U \odot AM_i^{inv}$. Specifically, we add an independent network branch $\psi_i$ using the trajectory of the agent $\tau_i$ to calculate the $\hat{\rho}^U$ and the $\rho_i^{inv}$. A common implementing method for encoding trajectory is to use the hidden state of the \textit{gated recurrent unit} (GRU) \cite{cho2014+GRU} $h_i$. Because the GRU takes $o_i^t$ and $a_i^t$ as the input recursively for all time-steps $t$, we assume that $h_i$ can appropriately represent $\tau_i$. Finally, since the UAS guarantees the capability of achieving the global optimality under the parameter-sharing constraint, the CGI loss is calculated by the \textit{mean squared error} (MSE) loss between $\rho^{inv}$ and $\rho$ for all agents with the network parameter $\boldsymbol{\psi} = (\psi_1, ..., \psi_n)$, and is written as:
\begin{equation}
    \mathcal{L}_{CGI}(\rho^{inv}_{\boldsymbol{\psi}} | h ) =\mathbb{E}_{\mathcal{D}} (\rho^{inv} - \rho)^2\ ,
    \label{equa-Loss-CGI-policy}
\end{equation}
where $\mathbb{E}_\mathcal{D}$ means to sample a batch of tuples $(\boldsymbol{\tau}, s, r)$ from replay buffer $\mathcal{D}$ and calculate expectation across the batch. For notational convenience, we also use $\mathbb{E}_\mathcal{D}$ for PPO-style on-policy methods without distinction. In that case, the replay buffer $\mathcal{D}$ only consists of the present episode.\par
Similarly, for value-based algorithms, since the action policy is usually calculated by the \textit{argmax} of Q values, the CGI loss is calculated by the MSE loss between $Q^{inv}$ and $Q$ for all agents and is written as:
\begin{equation}
    \mathcal{L}_{CGI}(Q^{inv}_{\boldsymbol{\psi}} | h ) =\mathbb{E}_{\mathcal{D}} (Q^{inv} - Q)^2\ .
    \label{equa-Loss-CGI-value}
\end{equation}

\begin{figure*}
\centering
\includegraphics[width=0.99\textwidth]{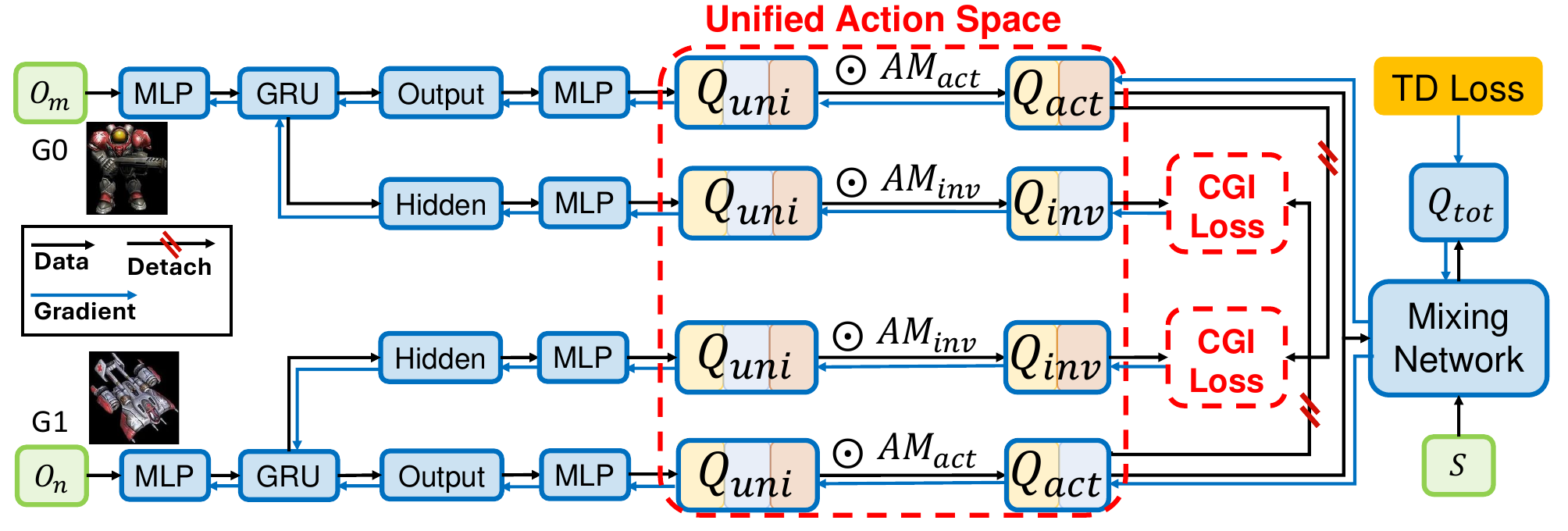}
\caption{An overall framework of U-QMIX. The overall network consists of two parts: Q network $\boldsymbol{\theta}$ and predictor network $\boldsymbol{\psi}$. The Q network includes the local Q network $\boldsymbol{\theta}_i$ and the mixing network $\boldsymbol{\theta}_M$. The local Q network $\boldsymbol{\theta}_i$ and the predictor network $\boldsymbol{\psi}$ are shared by all agents. They take their corresponding inputs and generate the UAS Q values $\boldsymbol{Q}_{uni}$. Then the $\boldsymbol{Q}_{uni}$ are masked by different available-action-masks $AM$ to generate different Q values for calculating different losses. The mixing network $\boldsymbol{\theta}_M$ is a global network to compute the $Q_{tot}$ during training.}
\label{fig-framework-U-QMIX}
\end{figure*}

\subsection{Practical Algorithms: U-MAPPO and U-QMIX}
In this section, we design a policy-based algorithm U-MAPPO and a value-based algorithm U-QMIX as examples to prove the efficiency of the UAS module and CGI loss. As is mentioned in section \ref{Preliminaries}, the different parts of network parameters are notated with $\theta$, $\phi$, and $\psi$. The bold fonts of $\boldsymbol{\theta}$, $\boldsymbol{\phi}$, and $\boldsymbol{\psi}$ indicate the usage of global parameter-sharing in the UAS optimization.

\subsubsection{U-MAPPO}
In order to implement the UAS based on policy-based methods, we combine the UAS with the MAPPO\cite{mappo+yu2021surprising} to form the U-MAPPO algorithm. The detailed network structure is illustrated in Fig. \ref{fig-framework-U-MAPPO}, and a pseudo-code is given in Algorithm \ref{algo-U-MAPPO}. \par
The overall network consists of three different sub-networks: the actor network $\boldsymbol{\theta} (a | \tau) = (\theta_1(a_1 | \tau_1), ..., \theta_n(a_n | \tau_n))$, the predictor network $\boldsymbol{\psi} (\rho^{inv} | h) = (\psi_1(\rho^{inv}_1 | h_1), ..., \psi_n(\rho^{inv}_n | h_n))$ and the critic network $\boldsymbol{\phi} (V | s)$. These parameters are responsible for the joint action policy $\pi_{\boldsymbol{\theta}} = (\pi_{\theta_1}, ..., \pi_{\theta_n})$, the joint inverse policy $\rho^{inv}_{\boldsymbol{\psi}} = (\rho^{inv}_{\psi_1}, ..., \rho^{inv}_{\psi_n})$ and the global value function $V_{\boldsymbol{\phi}}(s)$, respectively. \par
Following the implementation of MAPPO, the actor $\pi_{\boldsymbol{\theta}}$ is shared by all agents and is optimized following the loss function \ref{equa-Loss-U-MAPPO-actor}. It receives the current observation $o^t$, the last action $a^{t-1}$, and the last GRU hidden state $h^{t-1}$ to calculate the next action $a^t$. Due to the analysis in section \ref{CGI Loss}, we notate the observation-action trajectory as $\tau^t = (o^t, a^{t-1}, h^{t-1})$. The critic $V_{\boldsymbol{\phi}}$ is a mapping function $S \to \mathbb{R}$, and is centrally trained following the loss function \ref{equa-Loss-U-MAPPO-critic}. The additional predictor module $\rho^{inv}_{\boldsymbol{\psi}}$ is also shared by all agents for generating the inverse policy and is optimized following the CGI loss function \ref{equa-Loss-CGI-policy}.

\begin{equation}
\begin{aligned}
    \mathcal{L} (\pi_{\boldsymbol{\theta}}) &= \mathbb{E}_{\mathcal{D}}[\min(\frac{\pi_{\boldsymbol{\theta}}(a | \tau)}{\pi_{\boldsymbol{\theta}_{old}}(a | \tau)} A_{\pi_{\boldsymbol{\theta}}}(s,a), \\
    & clip(\frac{\pi_{\boldsymbol{\theta}}(a | \tau)}{\pi_{\boldsymbol{\theta}_{old}}(a | \tau)}, 1 \pm \epsilon_p ) A_{\pi_{\boldsymbol{\theta}}}(s,a) )] \\
    & + \lambda_{E}\mathbb{E}_{\mathcal{D}}[S(\pi_{\boldsymbol{\theta}}(\tau))] \ .
\end{aligned}
\label{equa-Loss-U-MAPPO-actor}
\end{equation}

\begin{equation}
\begin{aligned}
    \mathcal{L} (V_{\boldsymbol{\phi}}) &= \mathbb{E}_{\mathcal{D}}[\max((V_{\boldsymbol{\phi}}(s)- \hat R)^2, \\
    & ((clip(V_{\boldsymbol{\phi}}(s), V_{\boldsymbol{\phi}_{old}}(s) \pm \epsilon_v))- \hat R)^2)] \ .
\end{aligned}
\label{equa-Loss-U-MAPPO-critic}
\end{equation}

\begin{equation}
    \mathcal{L}_{U-MAPPO} = \mathcal{L} (\pi_{\boldsymbol{\theta}}) + \lambda_{V}\mathcal{L} (V_{\boldsymbol{\phi}}) + \lambda_{I}\mathcal{L}(\rho^{inv}_{\boldsymbol{\psi}})\ .
    \label{equa-Loss-U-MAPPO-all}
\end{equation}

The overall loss function of U-MAPPO is \ref{equa-Loss-U-MAPPO-all}. The \textit{clip} operator indicates clipping the value of the first variable to fit in the threshold interval determined by the second function with the hyper-parameter $\epsilon_p$ or $\epsilon_v$. $A_{\pi_{\boldsymbol{\theta}}}(s,a)$ is the advantage value computed with the GAE \cite{GAE-schulman2015high} method. $S(\pi_{\boldsymbol{\theta}}(\tau))$ is the policy entropy. $\hat R$ is the discounted reward-to-go. $\lambda_{E}$, $\lambda_{V}$, and $\lambda_{I}$ are the coefficients responsible for the entropy loss, value loss, and CGI loss, respectively.

\begin{algorithm}
\caption{U-MAPPO}
\label{algo-U-MAPPO}
\textbf{Input}: Learning rate $\alpha$, loss weights $\lambda_{V}$ and $\lambda_{I}$, number of groups $K$, number of agents $N$, max total time-steps $T^{TOT}$, max episodic time-steps $T^{EP}$, ppo-update times $U$.\\
\textbf{Initialize}: Network parameters $\boldsymbol{\theta}$, $\boldsymbol{\phi}$, and $\boldsymbol{\psi}$, total running time-step $t^{TOT}=0$.
\begin{algorithmic}[1] 
\While{$t^{TOT}\le T^{TOT}$}
\State Set episode step $t=0$, receive initial state and observation $(s^0, \boldsymbol{o}^0)$ from the environment.
\State Initialize episodic replay buffer $\mathcal{D} = \{\}$
\While{$t \le T^{EP}$ and \textbf{not} $terminated$}
\State Choose joint action $\boldsymbol{a}^t$ using actor network $\boldsymbol{\theta}$.
\State Interact with the environment using $\boldsymbol{a}^t$ and receive reward $r^{t}$, \textit{is\_terminated} and next state and observation $(s^{t+1}, \boldsymbol{o}^{t+1})$.
\State Collect a transition tuple at $t$ and update the replay buffer $\mathcal{D} = \mathcal{D} \cup \{(\boldsymbol{\tau}^{t}, s^{t}, r^{t})\}$.
\State $t=t+1$.
\EndWhile
\State Set ppo-update counter $u=0$.
\While{$u \le U$}
\State Calculate global value $V_{\boldsymbol{\phi}}$ and advantage $A_{\pi_{\boldsymbol{\theta}}}$ using critic network $\boldsymbol{\phi}$.
\State Calculate joint inverse policy $\rho^{inv}_{\boldsymbol{\psi}}$.
\State Calculate $\mathcal{L} (\pi_{\boldsymbol{\theta}})$ following (\ref{equa-Loss-U-MAPPO-actor}).
\State Calculate $\mathcal{L} (V_{\boldsymbol{\phi}})$ following (\ref{equa-Loss-U-MAPPO-critic}).
\State Calculate $\mathcal{L} (\rho^{inv}_{\boldsymbol{\psi}})$ following (\ref{equa-Loss-CGI-policy}).
\State Calculate total loss $\mathcal{L}_{U-MAPPO}$ following (\ref{equa-Loss-U-MAPPO-all}) and update $\boldsymbol{\theta}$, $\boldsymbol{\phi}$, and $\boldsymbol{\psi}$.
\State $u=u+1$.
\EndWhile
\State $t^{TOT}=t^{TOT}+t$.
\EndWhile
\end{algorithmic}
\end{algorithm}

\subsubsection{U-QMIX}
Similar to the U-MAPPO, we combine the UAS with the QMIX\cite{rashid2020+qmix-jmlr} to form the U-QMIX algorithm. The network structure is referred to in Fig. \ref{fig-framework-U-QMIX}, and the pseudo-code of U-QMIX is given in Algorithm \ref{algo-U-QMIX}. \par
In general, there are two parts of sub-networks optimized with different losses. The Q network includes the local Q network $\boldsymbol{\theta}_i (Q | \tau) = (\theta_1(Q_1 | \tau_1), ..., \theta_n(Q_n | \tau_n))$ and the mixing network $\boldsymbol{\theta}_M (Q_i, s)$. The predictor network is symbolized with $\boldsymbol{\psi} (Q^{inv} | h) = (\psi_1(Q^{inv}_1 | h_1), ..., \psi_n(Q^{inv}_n | h_n))$. \par
The local Q network $\boldsymbol{\theta}_i$ is shared by all agents. It uses the current observation $o_i^t$, the last action $a_i^{t-1}$ and the last GRU hidden state $h_i^{t-1}$ to calculate the next Q value $Q_{\boldsymbol{\theta}_i}^t$. Similarly, we notate the observation-action trajectory as $\tau_i^t = (o_i^t, a_i^{t-1}, h_i^{t-1})$. Mixing network $\boldsymbol{\theta}_M$ is a global network for value factorization. It takes all local $[Q_{\boldsymbol{\theta}_i}]$ as input, and mixes with the state $s$ to produce the $Q_{tot}$. The loss of $\boldsymbol{\theta}$ is calculated following \ref{equa-Loss-U-QMIX}, where the $Q^{tgt}_{tot}$ is the target Q value calculated by the target Q network $\boldsymbol{\theta}^{tgt}$. Predictor network $\psi$ is used to compute $Q^{inv}_{\boldsymbol{\psi}}$ and is optimized by the CGI loss function \ref{equa-Loss-CGI-value}. The total loss function of U-QMIX is written as \ref{equa-Loss-U-QMIX-all}. Similar to \ref{equa-Loss-U-MAPPO-all}, the $\lambda_{I}$ is the coefficient hyper-parameter responsible for the CGI loss. \par

\begin{equation}
\begin{aligned}
    &\mathcal{L} (Q_{\boldsymbol{\theta}})\ =\
    \mathbb{E}_\mathcal{D} [(y^{\boldsymbol{\theta}}-Q_{tot}(\boldsymbol{\tau},s;\boldsymbol{\theta}))^2]\ ,\\
    &y^{\boldsymbol{\theta}}\ =\ r+\gamma \ \max_{\boldsymbol{a}'} Q^{tgt}_{tot}(\boldsymbol{\tau}',s';\boldsymbol{\theta}^{tgt})\ ,
    \label{equa-Loss-U-QMIX}
\end{aligned}
\end{equation}

\begin{equation}
    \mathcal{L}_{U-QMIX} = \mathcal{L} (Q_{\boldsymbol{\theta}}) + \lambda_{I}\mathcal{L}(Q^{inv}_{\boldsymbol{\psi}})\ .
    \label{equa-Loss-U-QMIX-all}
\end{equation}

\begin{algorithm}
\caption{U-QMIX}
\label{algo-U-QMIX}
\textbf{Input}: Learning rate $\alpha$, loss weight $\lambda_{I}$, number of groups $K$, number of agents $N$, max total time-steps $T^{TOT}$, max episodic time-steps $T^{EP}$, batch size $B$.\\
\textbf{Initialize}: Network parameters $\boldsymbol{\theta}$ and $\boldsymbol{\psi}$, replay buffer $\mathcal{D} = \{\}$, total running time-step $t^{TOT}=0$.
\begin{algorithmic}[1] 
\While{ $t^{TOT}\le T^{TOT}$}
\State Set episode step $t=0$, receive initial state and observation $(s^0, \boldsymbol{o}^0)$ from the environment.
\While{$t \le T^{EP}$ and \textbf{not} $terminated$}
\State Choose joint action $\boldsymbol{a}^t$ using agent network $\boldsymbol{\theta}_i$.
\State Interact with the environment using $\boldsymbol{a}^t$ and receive reward $r^{t}$, \textit{is\_terminated} and next state and observation $(s^{t+1}, \boldsymbol{o}^{t+1})$.
\State Collect a transition tuple at $t$ and update the replay buffer $\mathcal{D} = \mathcal{D} \cup \{(\boldsymbol{\tau}^{t}, s^{t}, r^{t})\}$.
\State $t=t+1$.
\EndWhile
\State Sample a random batch of $B$ episodes from $\mathcal{D}$.
\State Calculate $Q_{tot}$ using mixing network $\boldsymbol{\theta}_M$.
\State Calculate joint inverse Q value $Q^{inv}_{\boldsymbol{\psi}}$.
\State Calculate $\mathcal{L} (Q_{\boldsymbol{\theta}})$ following (\ref{equa-Loss-U-QMIX}).
\State Calculate $\mathcal{L} (Q^{inv}_{\boldsymbol{\psi}})$ following (\ref{equa-Loss-CGI-value}).
\State Calculate total loss $\mathcal{L}_{U-QMIX}$ following (\ref{equa-Loss-U-QMIX-all}) and update $\boldsymbol{\theta}$ and $\boldsymbol{\psi}$.
\State $t^{TOT}=t^{TOT}+t$.
\EndWhile
\end{algorithmic}
\end{algorithm}

\subsubsection{Summary of Hyper-parameters}
Commonly, we choose Adam \cite{kingma2014adam} as the optimizer. The maximum time-steps of one episode is set to be 200. The reward discounting factor $\gamma$ is 0.99. We use the latest version 4.10 of the StarcraftII game on the Linux platform to perform experiments. \par
For U-MAPPO method, the learning rate is 5e-4 and the total training step is 10M. We set the threshold interval to be $\epsilon_p = \epsilon_v = 0.2$, and the loss coefficients to be $\lambda_{E} = 0.01$, $\lambda_{V} = 1$ and $\lambda_{I} = 0.8$. The Table \ref{table-hyper-parameters-U-MAPPO} summarizes the U-MAPPO hyper-parameters mentioned above. \par
For U-QMIX, the learning rate is 3e-4 and the total training step is 5M. The learning rate is scheduled to decay by multiplying the factor 0.5 every 50,000 episodes (averagely about 2M-3.5M steps). The $\epsilon$ of the $\epsilon -greedy$ action selecting policy starts at 1.0, ends at 0.05 and linearly declines for 50,000 steps. The size of the memory buffer $\mathcal{D}$ is 5,000 and the batch size is 32. The CGI loss coefficient is set to be $\lambda_{I} = 0.06$. The Table \ref{table-hyper-parameters-U-QMIX} summarizes the U-QMIX hyper-parameters mentioned above. \par

\begin{table}
\centering
\caption{The hyper-parameters of U-MAPPO.}
\label{table-hyper-parameters-U-MAPPO}
\begin{tabular}{*{4}{c}}
\toprule
Hyper\_  &  \multirow{2}*{Value} & Hyper\_  &  \multirow{2}*{Value} \\
parameters &  & parameters &   \\
\midrule
learning rate & 5e-4 & training step & 10M\\
$\epsilon_p$ & 0.2 & $\epsilon_v$ & 0.2 \\
$\lambda_{E}$ & 0.01 & $\lambda_{V}$ & 1 \\
$\lambda_{I}$ & 0.8 &  &  \\
\bottomrule
\end{tabular}
\end{table}

\begin{table}
\centering
\caption{The hyper-parameters of U-QMIX.}
\label{table-hyper-parameters-U-QMIX}
\begin{tabular}{*{4}{c}}
\toprule
Hyper\_  &  \multirow{2}*{Value} & Hyper\_  &  \multirow{2}*{Value} \\
parameters &  & parameters &   \\
\midrule
learning rate & 3e-4 & training step & 5M \\
lr\_decay factor & 0.5 & $\epsilon$ & 1.0 $\to$ 0.05\\
$\lambda_{I}$ & 0.06 &    &  \\
\bottomrule
\end{tabular}
\end{table}

\section{Experiments and Results}\label{Experiments and Results}
\subsection{Experimental Details}
We use 5 maps mentioned in \cite{GHQ-yu2024ghq} and the MMM2 map from the original map set to conduct our experiments. Detailed map information can be seen in Table \ref{table-SMAC-new-maps}. As we have introduced in section \ref{Relation-sec3.3} and \ref{UAS}, the Marine is an attacking unit and its action space consists of $A_s$ and $A_e$, while the Medivac is a supporting unit and its action space consists of $A_s$ and $A_a$. \par
For all agents, the dim of \textit{self} action $|A_s|$ is 6. Action ID 0 is the \textit{null} action only available for dead agents. Action ID 1 is the \textit{stop} action, and ID $ \{ 2, 3, 4, 5 \} $ are \textit{moving} actions, indicating moving up, down, left, and right. The  \textit{self} actions $A_s$ are available for all agents. The dim of \textit{ally} action $|A_a|$ equals the number of \textit{allies}, while the dim of \textit{enemy} action $|A_e|$ equals the number of \textit{enemies}. \par
We use the default global reward function of SMAC environment. Agents are rewarded when dealing \textit{damage} to the enemies, \textit{killing} enemies, and \textit{winning} the game. The \textit{damage} reward equals the absolute damage value dealt to the enemies. The \textit{killing} reward is a constant 10 for every enemy kill. And the \textit{winning} reward is a constant 200 given at the terminal time-step if the agents win the enemies. \par
We use the traditional winning-rate \textit{(WR)} as the measuring criterion. \textit{WR} is the probability of MARL agents eliminating all enemies and winning the game, and we use the average value of \textit{WR} over 32 testing episodes per every 10,000 training steps (about 1,000 training episodes). We take 5 rounds of individual experiments with different random seeds. Then, the curves of \textit{WR} changing over time are plotted with the p-value being 0.05.

\begin{table}
\centering
\caption{Asymmetric Heterogeneous SMAC Maps.}
\label{table-SMAC-new-maps}
\setlength{\tabcolsep}{1.8mm}{
\begin{tabular}{*{5}{c}}
\toprule
\multirow{2}*{Map Name} &  Ally\_  &  Ally\_  &   Enemy\_  & \multirow{2}*{Difficulty} \\
 &  Marines  &  Medivacs  &  Marines  & \\
\midrule
6m2m\_15m & 6 & 2 & 15 & Easy\\
6m2m\_16m & 6 & 2 & 16 & Medium\\
8m3m\_21m & 8 & 3 & 21 & Medium\\
8m4m\_23m & 8 & 4 & 23 & Hard\\
12m4m\_30m & 12 & 4 & 30 & Ex-Hard\\
\bottomrule
\end{tabular}
}
\end{table}

\subsection{Comparison Algorithms}
For policy-based algorithm comparison against the U-MAPPO, we choose COMA \cite{foerster2018+coma}, MAPPO \cite{mappo+yu2021surprising}, HAPPO \cite{happo+kuba2021trust} and MAT \cite{MAT-wen2022multi} as baselines. For value-based algorithm comparison against the U-QMIX, vanilla QMIX \cite{rashid2020+qmix-jmlr}, ROMA \cite{wang2020roma}, RODE \cite{wang2020rode}, CDS \cite{li2021+CDS}, ASN \cite{ASN-wang2019action}, UNMAS \cite{UNMAS-chai2023unmas} and GHQ \cite{GHQ-yu2024ghq} are used to conduct experiments. \par
In general, all of the aforementioned policy-based algorithms are actor-critic algorithms using the ``centralized critic decentralized actor'' (CCDA) architecture, which means that the state $s$ is only available for the critic network to estimate a proper value function $V(s)$, and the actor network is only capable of using the observation-action trajectory $\tau$ to generate the policy function $\pi(a|\tau)$. COMA and MAPPO use a shared actor network for all agents across heterogeneous agent types. HAPPO uses independent actor network parameters and proposes a theoretically monotonic policy-improving architecture with a sequential decision-making scheme. MAT further modifies the network architecture of HAPPO into the Transformer-style, and considers the MARL problem as a sequential decision-making problem to solve. \par
QMIX is a fundamental value-based MARL algorithm. Its introduction of monotonic value factorization architecture has been proven to be powerful yet efficient in SMAC environment. RODE and ROMA are both \textit{role-based} algorithms, which learn and apply role policies end to end. The default training step of ROMA is 20M, and thus it generally performs badly within our 5M training-step restriction. In RODE, several key hyper-parameters define the generation and usage of role policies, which require extra fine-tuning and restrict the performance of RODE. CDS introduces diversity into the optimization and representation of QMIX to overcome the shortage of global parameter-sharing. ASN considers the action semantics of different agents and divides the network to represent and leverage the semantics. UNMAS is capable of adapting to changes in the number and size of MAS with the self-weighting mixing network. GHQ divides agents into different groups using the IOG method and optimizes the networks of these groups with hybrid factorization and IGMI loss. \par

\begin{figure*}
    \centering
	\includegraphics[width=0.99\textwidth]{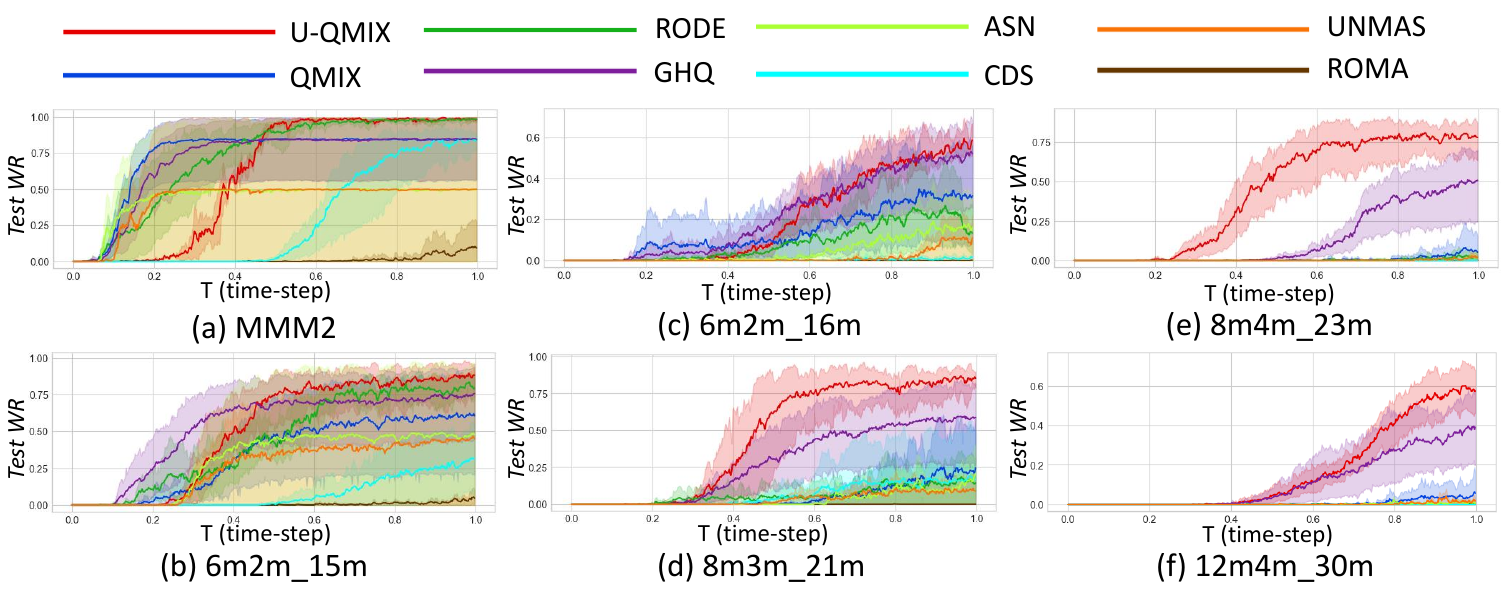}
	\caption{Results of Value-based Algorithms Comparison.}
	\label{fig-All-value_based}
\end{figure*}

\subsection{Comparison Results}
\begin{figure*}
    \centering
	\includegraphics[width=0.99\textwidth]{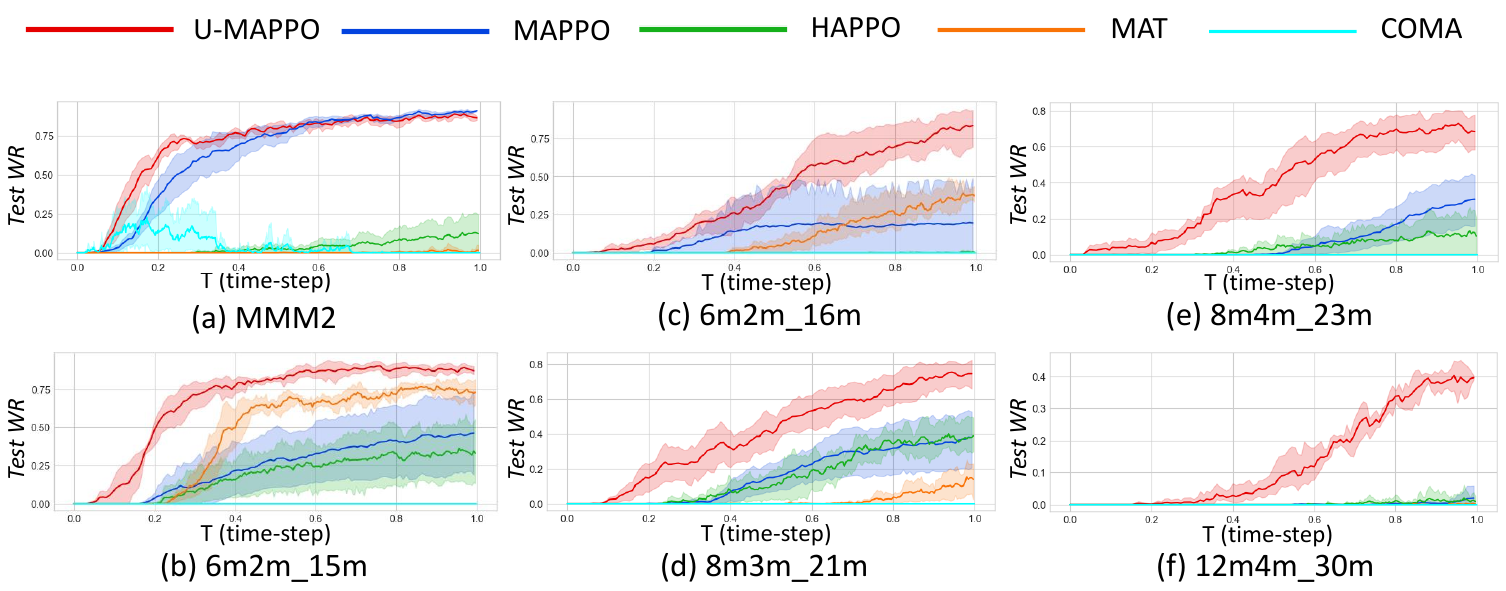}
	\caption{Results of Policy-based Algorithms Comparison.}
	\label{fig-All-policy_based}
\end{figure*}

The results of value-based algorithms are shown in section \ref{Value-based Algorithms Comparison}, Fig. \ref{fig-All-value_based}, and Table \ref{table-value-based}. The results of policy-based algorithms are shown in section \ref{Policy-based Algorithms Comparison}, Fig. \ref{fig-All-policy_based}, and Table \ref{table-policy-based}. The figures show the \textit{WR} curves changing over time and the tables show the final \textit{WR} values followed by the standard deviations. The results are averaged across 5 individual tests.\par
We criticize the performance of all algorithms from 3 aspects and the analysis is given below. Generally, all comparison algorithms are not specifically modified to deal with the global parameter-sharing dilemma in heterogeneous MARL problems, therefore they fail to achieve high \textit{WR} with small variance.

\subsubsection{Results of Value-based Algorithms Comparison}\label{Value-based Algorithms Comparison}

\begin{table*}
\centering
\caption{Results of Value-based Algorithms Comparison.}
\label{table-value-based}
\centering
\setlength{\tabcolsep}{1.7mm}{
\begin{tabular}{c|c|ccccccc}
\toprule
Map Name & U-QMIX & QMIX & GHQ & ROMA & RODE & CDS & ASN & UNMAS\\
\midrule
MMM2 & $\textbf{0.99}_{(0.03)}$ & $0.85_{(0.35)}$ & $0.85_{(0.35)}$ & $0.16_{(0.46)}$ & $\textbf{1.00}_{(0.01)}$ & $0.90_{(0.06)}$ & $0.49_{(0.48)}$ & $0.50_{(0.50)}$ \\
6m2m\_15m & $\textbf{0.88}_{(0.13)}$ & $0.68_{(0.31)}$ & $0.74_{(0.27)}$ & $0.18_{(0.21)}$ & $0.86_{(0.09)}$ & $0.60_{(0.12)}$ & $0.46_{(0.48)}$ & $0.41_{(0.40)}$\\
6m2m\_16m & $\textbf{0.58}_{(0.14)}$ & $0.33_{(0.23)}$ & $0.55_{(0.22)}$ & $0.02_{(0.03)}$ & $0.18_{(0.20)}$ & $0.08_{(0.08)}$ & $0.23_{(0.20)}$ & $0.14_{(0.16)}$ \\
8m3m\_21m & $\textbf{0.84}_{(0.13)}$ & $0.30_{(0.24)}$ & $0.66_{(0.28)}$ & $0.04_{(0.06)}$ & $0.16_{(0.25)}$ & $0.21_{(0.37)}$ & $0.23_{(0.21)}$ & $0.14_{(0.13)}$ \\
8m4m\_23m & $\textbf{0.81}_{(0.12)}$ & $0.13_{(0.20)}$ & $0.54_{(0.25)}$ & $0.00_{(0.00)}$ & $0.02_{(0.04)}$ & $0.07_{(0.11)}$ & $0.02_{(0.03)}$ & $0.02_{(0.02)}$ \\
12m4m\_30m & $\textbf{0.60}_{(0.15)}$ & $0.02_{(0.04)}$ & $0.40_{(0.27)}$ & $0.00_{(0.00)}$ & $0.01_{(0.01)}$ & $0.00_{(0.00)}$ & $0.02_{(0.02)}$ & $0.00_{(0.00)}$ \\
\bottomrule
\end{tabular}
}
\end{table*}

(1) \textit{Easy: we test all algorithms on the original asymmetric heterogeneous map MMM2.} The results are shown in Fig. \ref{fig-All-value_based} (a). U-QMIX and RODE converge to 1.0 at about 3M steps. The \textit{WR} curve of RODE increases faster than U-QMIX before 2M steps but suffers from high variance. QMIX and GHQ converge to about 0.8 at 2M steps. CDS also converges to 0.8 but requires 5M training steps. The \textit{WR} curve of ASN and UNMAS can only converge to 0.5 with high variance. ROMA is only capable of acquiring 0.1 at 5M training steps. \par
(2) \textit{Medium: we scale up ally units and increase enemy units for balancing.} In Fig. \ref{fig-All-value_based} (b), (c), (d), and (e), the number of ``Marines+Medivacs'' increases from ``6+2'' to ``8+3'' and ``8+4'', while the number of enemy Marines increases from 15 to 23. As is concluded in \cite{GHQ-yu2024ghq}, the power of 1 Medivac is roughly equal to 3.5 Marines. Therefore, the duels in these 4 maps are basically fair and the total amount of entities in the maps is also not too high.\par
Comparing the results in (a) and (b), it is easy to figure out that the difficulty of these two maps is similar, since all algorithms can acquire similar performance in these two maps. In (b), U-QMIX reaches the best \textit{WR} 0.9, and RODE reach the second \textit{WR} 0.8 with lower increasing rate. QMIX, ASN, UNMAS, and CDS end with about 0.6, 0.45, 0.4, and 0.3, respectively. In the first 2M steps, GHQ grows faster than all other algorithms, but converges at \textit{WR} 0.7 at about 3M steps. ROMA fails to learn effective policy in (b).\par
In (c), the increased 1 enemy Marine dramatically decreases the \textit{WR} of all algorithms. Only U-QMIX and GHQ are capable of reaching the \textit{WR} 0.5. QMIX, RODE, ASN, and UNMAS end with the \textit{WR} of about 0.3, 0.14, 0.15, and 0.13 separately. ROMA and CDS are both unable to learn any effective policy. In (d), we scale up allies with 1 Medivac and 2 Marines, and increase 5 enemy Marines for balancing. After 5M training steps, U-QMIX ends with the \textit{WR} 0.8 with relatively low variance. GHQ attains the \textit{WR} 0.6 with high variance. None of the other algorithms can reach the \textit{WR} higher than 0.25. Similarly, in (e), U-QMIX ends with the \textit{WR} 0.8 with low variance, while GHQ can only reach the \textit{WR} 0.5 with high variance. Still, none of the other algorithms can reach the \textit{WR} higher than 0.1. The above results significantly prove the effectiveness of U-QMIX against other baseline value-based methods in small-scale and medium-scale asymmetric heterogeneous SMAC maps.\par

\begin{table*}
\centering
\caption{Results of Policy-based Algorithms Comparison.}
\label{table-policy-based}
\centering
\begin{tabular}{c|c|cccc}
\toprule
Map Name & U-MAPPO & MAPPO & HAPPO & MAT & COMA \\
\midrule
MMM2 & $0.85_{(0.09)}$ & $\textbf{0.93}_{(0.02)}$ & $0.11_{(0.13)}$ & $0.02_{(0.03)}$ & $0.00_{(0.00)}$ \\
6m2m\_15m & $\textbf{0.85}_{(0.04)}$ & $0.47_{(0.38)}$ & $0.29_{(0.20)}$ & $0.76_{(0.09)}$ & $0.00_{(0.00)}$ \\
6m2m\_16m & $\textbf{0.85}_{(0.12)}$ & $0.19_{(0.33)}$ & $0.00_{(0.00)}$ & $0.35_{(0.11)}$ & $0.00_{(0.00)}$ \\
8m3m\_21m & $\textbf{0.83}_{(0.00)}$ & $0.38_{(0.24)}$ & $0.41_{(0.12)}$ & $0.12_{(0.12)}$ & $0.07_{(0.01)}$ \\
8m4m\_23m & $\textbf{0.74}_{(0.15)}$ & $0.31_{(0.20)}$ & $0.09_{(0.11)}$ & $0.00_{(0.00)}$ & $0.00_{(0.00)}$ \\
12m4m\_30m & $\textbf{0.41}_{(0.09)}$ & $0.02_{(0.05)}$ & $0.00_{(0.00)}$ & $0.00_{(0.00)}$ & $0.00_{(0.00)}$ \\
\bottomrule
\end{tabular}
\end{table*}

(3) \textit{Hard: we double all units of both sides simultaneously.} In Fig. \ref{fig-All-value_based} (b) and (f), the number of all entities is doubled. Theoretically, the optimal policies of maps (b) and (f) are similar using the ``divide and conquer'' method. However, most of the baseline algorithms fail to deal with the complex combination of scalability and heterogeneity. On the one hand, as we have analyzed before, U-QMIX, RODE, GHQ, and QMIX can achieve the \textit{WR} greater than 0.5 with relatively low variance in (b). Nevertheless, on the other hand, only U-QMIX and GHQ are capable of achieving significant \textit{WR} in (f). U-QMIX not only wins GHQ by higher \textit{WR} and lower variance, but also reaches \textit{WR} 0.6, which is very close to its performance in (b). This result indicates the superiority of U-QMIX in dealing with heterogeneous scalability problems.\par

\subsubsection{Results of Policy-based Algorithms Comparison}\label{Policy-based Algorithms Comparison}
Because the action space is discrete in SMAC, value-based algorithms generally perform better than policy-based algorithms \cite{rashid2020+qmix-jmlr,riit+hu2021rethinking,mappo+yu2021surprising,GHQ-yu2024ghq}, which means that value-based methods usually require fewer training-steps to achieve similar \textit{WR} than policy-based methods. Therefore, for a fair comparison and better performance, the training-step for policy-based algorithms is 10M. \par
(1) \textit{Easy: we test all algorithms on the original asymmetric heterogeneous map MMM2.} In Fig. \ref{fig-All-policy_based} (a), U-MAPPO and MAPPO converged to \textit{WR} 0.9 at about 7M training-steps. COMA acquires a little \textit{WR} before the first 3M steps and then fails to learn effective policy. HAPPO reaches \textit{WR} 0.1 at the end of training, while MAT fails to learn any policy.\par
(2) \textit{Medium: we scale up ally units and increase enemy units for balancing.} In Fig. \ref{fig-All-policy_based} (b), (c), (d), and (e), U-MAPPO impressively outperforms all other baseline methods with high \textit{WR} and low variance. MAT becomes the second method in (b) and (c), but still fails in (d) and (e), indicating that the sequential decision-making architecture requires modification in heterogeneous and large-scale MARL problems. MAPPO is capable of acquiring a \textit{WR} of 0.2 to 0.4 with a relatively large variance. HAPPO completely fails in (c) but has similar performance with MAPPO in (b), (d), and (e). COMA fails in all of the 4 maps.\par
(3) \textit{Hard: we double all units of both sides simultaneously.} Although most of the baseline algorithms achieve good performance in (b), only U-MAPPO can reach the \textit{WR} 0.4 with low variance. As we have mentioned before, (f) is the most complex and extremely hard map that combines the complexity of heterogeneity and scalability. The outstanding performance of U-MAPPO indicates the effectiveness of the UAS and CGI loss. \par

\subsection{Ablation Study}

\begin{figure}
    \centering
    \includegraphics[width=0.99\columnwidth]{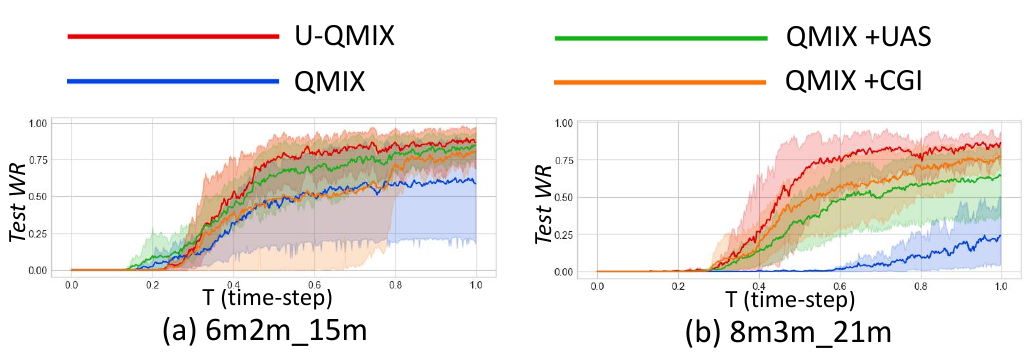}
    \caption{Results for ablation tests about UAS and CGI on value-based algorithm.}
    \label{fig-ablation-value_based}
\end{figure}

\begin{figure}
    \centering
    \includegraphics[width=0.99\columnwidth]{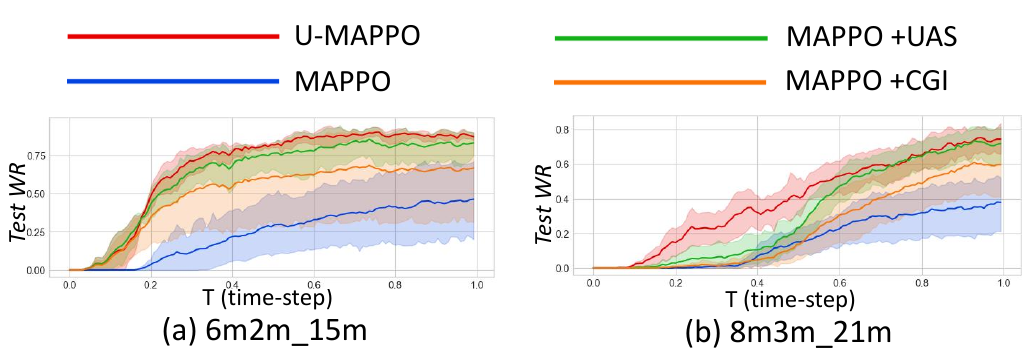}
    \caption{Results for ablation tests about UAS and CGI on policy-based algorithm.}
    \label{fig-ablation-policy_based}
\end{figure}

In order to determine the effectiveness of the UAS and CGI loss, we take ablation tests of the two components in (2) 6m2m\_15m, and (b) 8m3m\_21m. For value-based methods, QMIX is the baseline algorithm while ``QMIX+UAS'' and ``QMIX+CGI'' are the ablation groups. Similarly, for policy-based methods, MAPPO is the baseline algorithm while ``MAPPO+UAS'' and ``MAPPO+CGI'' are the ablation groups. The results are shown in Fig. \ref{fig-ablation-value_based} and \ref{fig-ablation-policy_based}, respectively. \par
In general, the introduction of UAS can improve the performance and reduce the varience, and the performance of CGI loss is unstable. In Fig. \ref{fig-ablation-value_based} (a), QMIX+UAS significantly outperforms QMIX with higher \textit{WR} and lower variance, indicating the effectiveness of distinguishing different action semantics with UAS. QMIX+CGI can still outperform QMIX in the last 1M training-steps, however,  it suffers from high variance. In Fig. \ref{fig-ablation-value_based} (b), however, QMIX+CGI acquires higher \textit{WR} than QMIX+UAS. In Fig. \ref{fig-ablation-policy_based} (a), the performances of MAPPO+UAS and MAPPO+CGI are similar to the performances in Fig. \ref{fig-ablation-value_based} (a).  In Fig. \ref{fig-ablation-policy_based} (b), the performances of MAPPO+UAS and MAPPO+CGI are similar in the first 5M training-steps, but MAPPO+UAS then outperforms MAPPO+CGI and reaches a similar \textit{WR} with the U-MAPPO.\par
One possible reason for the previous phenomenon is that the two diverse optimizing objects of the CGI loss against the original MARL loss weaken the stability of the learning process. Using UAS makes the backward propagation process more accurate and therefore improves the \textit{WR}. The combination of the UAS and CGI loss brings a larger network capacity and thus leads to the best performance of U-QMIX and U-MAPPO. In addition, for map (b) with larger scale and heterogeneity, the usage of CGI loss brings more increasing on the \textit{WR} in contrast with the ``UAS-only'' method, indicating the necessity of improving the cooperation and correlation between different groups in complex heterogeneous MARL problems.

\section{Conclusion}\label{Conclusion}
This paper focuses on the global parameter-sharing in physical heterogeneous MARL problems. Our core motivation is to design a proper global parameter-sharing method that can not only cooperate well but also avoid bad local optimality. Inspired by the term \textit{action semantics}, we introduce the \textit{unified action space} (UAS) as a solution for distinguishing the different action semantics of heterogeneous agents. The usage of UAS is capable of acquiring better performance than the baseline method since the accurate action semantics improve the optimization process. \par
To further improve the cooperation between heterogeneous groups, we introduce the \textit{cross group inverse} (CGI) loss which uses the trajectory of one group to predict the policy or Q value of other groups. This method avoids the calculation of mutual information, which must be taken between two groups, and therefore dramatically reduces the overall computation cost. We combine the UAS and CGI with standard value-based and policy-based methods to generate the U-QMIX and U-MAPPO methods.\par
Comparing and ablation experiments are conducted in physical heterogeneous SMAC maps to prove the effectiveness of U-QMIX and U-MAPPO. The results show that U-QMIX and U-MAPPO outperform other state-of-the-art algorithms with higher winning-rate and lower variance. \par
In the future, we will try to focus on the other type of heterogeneity, behavioral heterogeneity. We hope our work could be helpful for further study in the heterogeneous MARL problems. In addition, we are open to discover the complex combination of communication or scalability with heterogeneity.

\section*{Statements and Declarations}
\begin{itemize}
\item Funding: No funding was received to assist with the preparation of this manuscript.
\item Competing interests: The authors have no competing interests to declare that are relevant to the content of this article.
\item Ethics approval: This article does not involve any ethical problem which needs approval.
\item Consent to participate: All authors have seen and approved the final version of the manuscript being submitted.
\item Consent for publication: All authors warrant that the article is our original work, hasn't received prior publication, and isn't under consideration for publication elsewhere. A preprint version of our manuscript has been submitted to arXiv, and the page is xxx.
\item Availability of data and materials: The datasets generated during and/or analyzed during the current study are available from the corresponding author on reasonable request.
\item Code availability: The codes for this article are available from the corresponding author on reasonable request.
\item Authors' contributions: Conceptualization: [Xiaoyang Yu, Kai Lv, Shuo Wang]; Methodology: [Xiaoyang Yu, Shuo Wang]; Formal analysis and investigation: [Xiaoyang Yu]; Writing - original draft preparation: [Xiaoyang Yu]; Writing - review and editing: [Xiaoyang Yu, Youfang Lin, Shuo Wang, Sheng Han, Kai Lv]; Funding acquisition: [Youfang Lin, Sheng Han, Kai Lv]; Resources: [Youfang Lin, Sheng Han, Kai Lv]; Supervision: [Youfang Lin, Sheng Han, Kai Lv].
\end{itemize}

\makeatletter
\renewcommand\@biblabel[1]{#1.}
\makeatother
\bibliography{UAS}

\begin{thebibliography}{54}
\providecommand{\natexlab}[1]{#1}
\providecommand{\url}[1]{{#1}}
\providecommand{\urlprefix}{URL }
\providecommand{\doi}[1]{\url{https://doi.org/#1}}
\providecommand{\eprint}[2][]{\url{#2}}
 \bibcommenthead

\bibitem[{Oroojlooy and Hajinezhad(2022)}]{APPI-oroojlooy2022review}
Oroojlooy A, Hajinezhad D (2022) A review of cooperative multi-agent deep reinforcement learning. Applied Intelligence pp 1--46

\bibitem[{Gronauer and Diepold(2022)}]{survey+gronauer2022multi}
Gronauer S, Diepold K (2022) Multi-agent deep reinforcement learning: a survey. Artificial Intelligence Review 55(2):895--943

\bibitem[{Zhu et~al(2024)Zhu, Dastani, and Wang}]{zhu2024survey}
Zhu C, Dastani M, Wang S (2024) A survey of multi-agent deep reinforcement learning with communication. Autonomous Agents and Multi-Agent Systems 38(1):4

\bibitem[{Hernandez-Leal et~al(2019)Hernandez-Leal, Kartal, and Taylor}]{hernandez2019survey}
Hernandez-Leal P, Kartal B, Taylor ME (2019) A survey and critique of multiagent deep reinforcement learning. Autonomous Agents and Multi-Agent Systems 33(6):750--797

\bibitem[{Bal{\'a}zs et~al(2023)Bal{\'a}zs, Vicsek, Somorjai, Nepusz, and V{\'a}s{\'a}rhelyi}]{balazs2023decentralized}
Bal{\'a}zs B, Vicsek T, Somorjai G, et~al (2023) Decentralized traffic management of autonomous drones. arXiv preprint arXiv:231211207

\bibitem[{Sun et~al(2023)Sun, Wu, Shi, Yu, Gao, Pei, Yang, Piao, and Hou}]{NCA-sun2023multi}
Sun Z, Wu H, Shi Y, et~al (2023) Multi-agent air combat with two-stage graph-attention communication. Neural Computing and Applications 35(27):19765--19781

\bibitem[{Amorosa et~al(2023)Amorosa, Skocaj, Verdone, and G{\"u}nd{\"u}z}]{amorosa2023multi}
Amorosa LM, Skocaj M, Verdone R, et~al (2023) Multi-agent reinforcement learning for power control in wireless networks via adaptive graphs. arXiv preprint arXiv:231115858

\bibitem[{Holand et~al(2024)Holand, Homer, Storrer, Khandeker, Muhlon, Patel, Vainqueur, Antaki, Cooke, Wilson et~al}]{holand2024battery}
Holand E, Homer J, Storrer A, et~al (2024) Battery-swapping multi-agent system for sustained operation of large planetary fleets. arXiv preprint arXiv:240108497

\bibitem[{Allahham et~al(2022)Allahham, Abdellatif, Mhaisen, Mohamed, Erbad, and Guizani}]{allahham2022multi}
Allahham MS, Abdellatif AA, Mhaisen N, et~al (2022) Multi-agent reinforcement learning for network selection and resource allocation in heterogeneous multi-rat networks. IEEE Transactions on Cognitive Communications and Networking 8(2):1287--1300

\bibitem[{Lu et~al(2023)Lu, Li, Li, and Xu}]{NCA-lu2023maddpg}
Lu K, Li RD, Li MC, et~al (2023) Maddpg-based joint optimization of task partitioning and computation resource allocation in mobile edge computing. Neural Computing and Applications 35(22):16559--16576

\bibitem[{Ma et~al(2023)Ma, Zhang, Li, and Xu}]{NCA-ma2023multi}
Ma C, Zhang J, Li Z, et~al (2023) Multi-agent deep reinforcement learning algorithm with trend consistency regularization for portfolio management. Neural Computing and Applications 35(9):6589--6601

\bibitem[{Huang et~al(2024)Huang, Li, Yao, and Chen}]{NCA-huang2024mgcrl}
Huang Z, Li F, Yao J, et~al (2024) Mgcrl: Multi-view graph convolution and multi-agent reinforcement learning for dialogue state tracking. Neural Computing and Applications 36(9):4829--4846

\bibitem[{Liu and Ding(2022)}]{traffic+liu2022distributed}
Liu B, Ding Z (2022) A distributed deep reinforcement learning method for traffic light control. Neurocomputing 490:390--399

\bibitem[{Zhao et~al(2024)Zhao, Lin, Zhang, Li, Zhou, and Sun}]{NCA-zhao2024mimic}
Zhao J, Lin J, Zhang X, et~al (2024) From mimic to counteract: a two-stage reinforcement learning algorithm for google research football. Neural Computing and Applications pp 1--17

\bibitem[{Samvelyan et~al(2019)Samvelyan, Rashid, De~Witt, Farquhar, Nardelli, Rudner, Hung, Torr, Foerster, and Whiteson}]{smac+samvelyan2019starcraft}
Samvelyan M, Rashid T, De~Witt CS, et~al (2019) The starcraft multi-agent challenge. arXiv preprint arXiv:190204043

\bibitem[{Clauset et~al(2009)Clauset, Shalizi, and Newman}]{clauset2009power}
Clauset A, Shalizi CR, Newman ME (2009) Power-law distributions in empirical data. SIAM review 51(4):661--703

\bibitem[{Wang et~al(2024)Wang, Wu, Hu, Wang, Lin, and Lv}]{wang2024what}
Wang S, Wu Z, Hu X, et~al (2024) What effects the generalization in visual reinforcement learning: Policy consistency with truncated return prediction. Proceedings of the AAAI conference on artificial intelligence 38(1)

\bibitem[{Wang et~al(2023)Wang, Wu, Hu, Lin, and Lv}]{wang2023skill}
Wang S, Wu Z, Hu X, et~al (2023) Skill-based hierarchical reinforcement learning for target visual navigation. IEEE Transactions on Multimedia

\bibitem[{Lv et~al(2023)Lv, Wang, Han, and Lin}]{lv2023spatially}
Lv K, Wang S, Han S, et~al (2023) Spatially-regularized features for vehicle re-identification: An explanation of where deep models should focus. IEEE Transactions on Intelligent Transportation Systems

\bibitem[{Lv et~al(2020)Lv, Sheng, Xiong, Li, and Zheng}]{lv2020pose}
Lv K, Sheng H, Xiong Z, et~al (2020) Pose-based view synthesis for vehicles: A perspective aware method. IEEE Transactions on Image Processing 29:5163--5174

\bibitem[{Yu et~al(2021)Yu, Liew, and Wang}]{wifi+yu2021multi}
Yu Y, Liew SC, Wang T (2021) Multi-agent deep reinforcement learning multiple access for heterogeneous wireless networks with imperfect channels. IEEE Transactions on Mobile Computing

\bibitem[{Xing(2024)}]{xing2024designing}
Xing F (2024) Designing heterogeneous llm agents for financial sentiment analysis. arXiv preprint arXiv:240105799

\bibitem[{Ivi{\'c}(2020)}]{robotic+ivic2020motion}
Ivi{\'c} S (2020) Motion control for autonomous heterogeneous multiagent area search in uncertain conditions. IEEE Transactions on Cybernetics

\bibitem[{Yoon et~al(2019)Yoon, Chen, Long, Zhang, Gahlawat, Lee, and Hovakimyan}]{robotic+yoon2019learning}
Yoon HJ, Chen H, Long K, et~al (2019) Learning to communicate: A machine learning framework for heterogeneous multi-agent robotic systems. AIAA Scitech 2019 Forum p 1456

\bibitem[{Bettini et~al(2023)Bettini, Shankar, and Prorok}]{Def-HetGPPO-bettini2023heterogeneous}
Bettini M, Shankar A, Prorok A (2023) Heterogeneous multi-robot reinforcement learning. Proceedings of the 2023 International Conference on Autonomous Agents and Multiagent Systems pp 1485--1494

\bibitem[{Wilson et~al(2022)Wilson, King, and Peterson}]{Def-wilson2022evolution}
Wilson RJ, King DW, Peterson GL (2022) Evolution of combined arms tactics in heterogeneous multi-agent teams. The International FLAIRS Conference Proceedings, 35

\bibitem[{Wilson(2022)}]{Def-wilson2022performance}
Wilson RJ (2022) Performance of heterogeneous multi-agent systems with applications in combined arms. Theses and Dissertations 5330

\bibitem[{Yu et~al(2024)Yu, Lin, Wang, Han, and Lv}]{GHQ-yu2024ghq}
Yu X, Lin Y, Wang X, et~al (2024) Ghq: grouped hybrid q-learning for cooperative heterogeneous multi-agent reinforcement learning. Complex \& Intelligent Systems pp 1--20

\bibitem[{Kuba et~al(2021)Kuba, Chen, Wen, Wen, Sun, Wang, and Yang}]{happo+kuba2021trust}
Kuba JG, Chen R, Wen M, et~al (2021) Trust region policy optimisation in multi-agent reinforcement learning. arXiv preprint arXiv:210911251

\bibitem[{Bono et~al(2018)Bono, Dibangoye, Matignon, Pereyron, and Simonin}]{mapg+bono2018cooperative}
Bono G, Dibangoye JS, Matignon L, et~al (2018) Cooperative multi-agent policy gradient. Joint European Conference on Machine Learning and Knowledge Discovery in Databases pp 459--476

\bibitem[{Rashid et~al(2020)Rashid, Samvelyan, De~Witt, Farquhar, Foerster, and Whiteson}]{rashid2020+qmix-jmlr}
Rashid T, Samvelyan M, De~Witt CS, et~al (2020) Monotonic value function factorisation for deep multi-agent reinforcement learning. The Journal of Machine Learning Research 21(1):7234--7284

\bibitem[{Yu et~al(2021)Yu, Velu, Vinitsky, Wang, Bayen, and Wu}]{mappo+yu2021surprising}
Yu C, Velu A, Vinitsky E, et~al (2021) The surprising effectiveness of ppo in cooperative, multi-agent games. arXiv preprint arXiv:210301955

\bibitem[{Dong et~al(2021)Dong, Wang, Liu, Han, and Zhang}]{homophily+dong2021birds}
Dong H, Wang T, Liu J, et~al (2021) Birds of a feather flock together: A close look at cooperation emergence via multi-agent rl. arXiv preprint arXiv:210411455

\bibitem[{Wang et~al(2020)Wang, Yang, Liu, Hao, Hao, Hu, Chen, Fan, and Gao}]{ASN-wang2019action}
Wang W, Yang T, Liu Y, et~al (2020) Action semantics network: Considering the effects of actions in multiagent systems. International Conference on Learning Representations

\bibitem[{Foerster et~al(2016)Foerster, Assael, De~Freitas, and Whiteson}]{ctde+foerster2016learning}
Foerster J, Assael IA, De~Freitas N, et~al (2016) Learning to communicate with deep multi-agent reinforcement learning. Advances in neural information processing systems 29

\bibitem[{Kraemer and Banerjee(2016)}]{ctde+kraemer2016multi}
Kraemer L, Banerjee B (2016) Multi-agent reinforcement learning as a rehearsal for decentralized planning. Neurocomputing 190:82--94

\bibitem[{Gupta et~al(2017)Gupta, Egorov, and Kochenderfer}]{ctde+gupta2017cooperative}
Gupta JK, Egorov M, Kochenderfer M (2017) Cooperative multi-agent control using deep reinforcement learning. International conference on autonomous agents and multiagent systems pp 66--83

\bibitem[{Foerster et~al(2018)Foerster, Farquhar, Afouras, Nardelli, and Whiteson}]{foerster2018+coma}
Foerster J, Farquhar G, Afouras T, et~al (2018) Counterfactual multi-agent policy gradients. Proceedings of the AAAI conference on artificial intelligence 32(1)

\bibitem[{Wen et~al(2022)Wen, Kuba, Lin, Zhang, Wen, Wang, and Yang}]{MAT-wen2022multi}
Wen M, Kuba J, Lin R, et~al (2022) Multi-agent reinforcement learning is a sequence modeling problem. Advances in Neural Information Processing Systems 35:16509--16521

\bibitem[{Son et~al(2019)Son, Kim, Kang, Hostallero, and Yi}]{son2019qtran}
Son K, Kim D, Kang WJ, et~al (2019) Qtran: Learning to factorize with transformation for cooperative multi-agent reinforcement learning. International conference on machine learning pp 5887--5896

\bibitem[{Sunehag et~al(2017)Sunehag, Lever, Gruslys, Czarnecki, Zambaldi, Jaderberg, Lanctot, Sonnerat, Leibo, Tuyls et~al}]{vdn+sunehag2017value}
Sunehag P, Lever G, Gruslys A, et~al (2017) Value-decomposition networks for cooperative multi-agent learning. arXiv preprint arXiv:170605296

\bibitem[{Wang et~al(2020{\natexlab{a}})Wang, Gupta, Mahajan, Peng, Whiteson, and Zhang}]{wang2020rode}
Wang T, Gupta T, Mahajan A, et~al (2020{\natexlab{a}}) Rode: Learning roles to decompose multi-agent tasks. arXiv preprint arXiv:201001523

\bibitem[{Wang et~al(2020{\natexlab{b}})Wang, Dong, Lesser, and Zhang}]{wang2020roma}
Wang T, Dong H, Lesser V, et~al (2020{\natexlab{b}}) Roma: Multi-agent reinforcement learning with emergent roles. arXiv preprint arXiv:200308039

\bibitem[{Li et~al(2021)Li, Wang, Wu, Zhao, Yang, and Zhang}]{li2021+CDS}
Li C, Wang T, Wu C, et~al (2021) Celebrating diversity in shared multi-agent reinforcement learning. Advances in Neural Information Processing Systems 34:3991--4002

\bibitem[{Chai et~al(2023)Chai, Li, Zhu, Zhao, Ma, Sun, and Ding}]{UNMAS-chai2023unmas}
Chai J, Li W, Zhu Y, et~al (2023) Unmas: Multiagent reinforcement learning for unshaped cooperative scenarios. IEEE Transactions on neural networks and learning systems 34(4):2093--2104

\bibitem[{Liu et~al(2024)Liu, Zhong, Hu, Fu, Fu, Chang, and Yang}]{HASAC-2024liumaximum}
Liu J, Zhong Y, Hu S, et~al (2024) Maximum entropy heterogeneous-agent reinforcement learning. The Twelfth International Conference on Learning Representations

\bibitem[{Guo et~al(2024)Guo, Shi, Yu, and Fan}]{SHPPO-guo2024heterogeneous}
Guo X, Shi D, Yu J, et~al (2024) Heterogeneous multi-agent reinforcement learning for zero-shot scalable collaboration. arXiv preprint arXiv:240403869

\bibitem[{Sahoo et~al(2024)Sahoo, Tripathi, Saha, and Mondal}]{FedMRL-sahoo2024fedmrl}
Sahoo P, Tripathi A, Saha S, et~al (2024) Fedmrl: Data heterogeneity aware federated multi-agent deep reinforcement learning for medical imaging. arXiv preprint arXiv:240705800

\bibitem[{Zhou et~al(2024)Zhou, Piao, Chi, Chen, and Li}]{HeR-DRL-zhou2024her}
Zhou X, Piao S, Chi W, et~al (2024) Her-drl: Heterogeneous relational deep reinforcement learning for decentralized multi-robot crowd navigation. arXiv preprint arXiv:240310083

\bibitem[{Oliehoek and Amato(2016)}]{DEC-POMDP+oliehoek2016concise}
Oliehoek FA, Amato C (2016) A concise introduction to decentralized POMDPs. Springer

\bibitem[{Cho et~al(2014)Cho, van Merrienboer, G{\"u}l{\c{c}}ehre, Bahdanau, Bougares, Schwenk, and Bengio}]{cho2014+GRU}
Cho K, van Merrienboer B, G{\"u}l{\c{c}}ehre {\c{C}}, et~al (2014) Learning phrase representations using rnn encoder-decoder for statistical machine translation. EMNLP

\bibitem[{Schulman et~al(2016)Schulman, Moritz, Levine, Jordan, and Abbeel}]{GAE-schulman2015high}
Schulman J, Moritz P, Levine S, et~al (2016) High-dimensional continuous control using generalized advantage estimation. Proceedings of the International Conference on Learning Representations (ICLR)

\bibitem[{Kingma and Ba(2014)}]{kingma2014adam}
Kingma DP, Ba J (2014) Adam: A method for stochastic optimization. arXiv preprint arXiv:14126980

\bibitem[{Hu et~al(2021)Hu, Jiang, Harding, Wu, and Liao}]{riit+hu2021rethinking}
Hu J, Jiang S, Harding SA, et~al (2021) Rethinking the implementation tricks and monotonicity constraint in cooperative multi-agent reinforcement learning. arXiv e-prints pp arXiv--2102

\end{thebibliography}

\end{document}